\documentclass[prb,twocolumn,showpacs,preprintnumbers,amsmath,amssymb,unsortedaddress]{revtex4-1}

\usepackage{graphicx}
\usepackage{dcolumn}
\usepackage{bm}

\begin{document}

\title{Theory of magnetic small-angle neutron scattering of two-phase ferromagnets}

\author{Dirk~Honecker}
\author{Andreas~Michels}
\email[Corresponding author. Electronic address: ]{andreas.michels@uni.lu}
\affiliation{Physics and Material Sciences Research Unit, University of Luxembourg, 162A~Avenue de la Fa{\"i}encerie, L-1511 Luxembourg, Grand Duchy of Luxembourg}

\date{\today}

\begin{abstract}
Based on micromagnetic theory we have derived analytical expressions for the magnetic small-angle neutron scattering (SANS) cross section of a two-phase particle-matrix-type ferromagnet. The approach---valid close to magnetic saturation---provides access to several features of the spin structure such as perturbing magnetic anisotropy and magnetostatic fields. Depending on the applied magnetic field and on the magnitude $H_p$ of the magnetic anisotropy field relative to the magnitude $\Delta M$ of the jump in the longitudinal magnetization at the particle-matrix interface, we observe a variety of angular anisotropies in the magnetic SANS cross section. In particular, the model explains the ``clover-leaf''-shaped angular anisotropy which was previously observed for several nanostructured magnetic materials, and it provides access to the magnetic interaction parameters such as the average exchange-stiffness constant. It is also shown that the ratio $H_p / \Delta M$ decisively determines the asymptotic power-law exponent and the range of spin-misalignment correlations.
\end{abstract}


\maketitle\

\section{Introduction}

Magnetic small-angle neutron scattering (SANS) is one of the most important techniques for microstructure determination in magnetism and magnetic materials research. The essential advantages of the SANS method may be summarized by stating that (i) SANS provides nanometer-scale information ($\sim 1 - 100 \, \mathrm{nm}$) from within the bulk of a sample and that (ii) SANS experiments can be conducted under rather flexible conditions and different sample environments (temperature, electric and magnetic field, pressure, neutron polarization, time-resolved data acquisition, etc.). For instance, SANS has previously been used for investigating the anisotropy of the structure factor and the relaxation dynamics of magnetic fluids, \cite{gazeau02,albi06,albi11} the internal structure and morphology of magnetic nanoparticles and nanowire arrays, \cite{thomson05,ijiri05,grigorieva07,krycka2010,gri11,disch2012} the vortex lattice of superconductors, \cite{forgan2011} or the spin distribution of nanocrystalline \cite{heinemann2000,michels00a,michels00b,michels03epl,herr04pss,loeff05,kohlbrecher05,michels2010epjb,dobrichprb2012,michels2012prb2,bickapl2013} and amorphous \cite{bracchi04,garcia04,cald05,mergia08} metals.

The scattering contrast in magnetic SANS arises from deviations of both the magnitude and orientation of the local magnetization vector field $\mathbf{M}(\mathbf{r})$ from its mean value. The phenomenological continuum theory of micromagnetics \cite{brown,aharonibook,kronfahn03} allows one to compute the magnetic microstructure of a material, i.e., the variation of the direction of $\mathbf{M}$ as a function of the position $\mathbf{r}$ and time. Since the elastic magnetic differential scattering cross section depends on the Fourier coefficients of the magnetization, it is, in principle, straightforward to use micromagnetic theory for predicting and modeling of magnetic SANS. \cite{erokhin2012prb,michels2012prb1,michels2013jmmm} However, due to the nonlinearity of the governing equations---Brown's equations of micromagnetics---it is rather difficult to obtain analytical closed-form expressions for the cross section and, indeed, such solutions are limited to the approach-to-saturation regime where the micromagnetic equations can be linearized.

Based on the pioneering work of Kronm\"uller {\it et al.} (Ref.~\onlinecite{kron63}) we have previously derived analytical expressions for the differential magnetic scattering cross section of ferromagnets with \emph{uniform} values of the saturation magnetization and the exchange interaction, but with a highly nonuniform magnetocrystalline and/or magnetoelastic anisotropy. \cite{jnist,jprb2001} Examples for such materials are nanocrystalline or cold-worked metals which exhibit nanoscale variations in the direction and/or magnitude of the magnetic anisotropy, leading to a highly inhomogeneous spin structure along with a strongly field-dependent magnetic SANS signal. As summarized in Ref.~\onlinecite{michels08rop}, analysis of experimental SANS data on nanocrystalline cobalt and nickel and on cold-worked coarse-grained nickel yields information on the strength and spatial structure of the magnetic anisotropy field, on the magnitude and field dependence of the magnetostatic field, on the exchange constant, and on the characteristic length of the spin misalignment.

It is the purpose of this article to extend the micromagnetic scattering approach to materials having \emph{nonuniform} saturation magnetization. Prominent examples are two-phase hard and soft magnetic nanocomposites, which are used e.g.\ in electronics devices, transformers or motors. \cite{suzuki06,gutfleisch2011} Here, the jump in the magnitude of the magnetization at the particle-matrix interface gives rise to a magnetostatic stray field which represents a dominating source of spin disorder.\cite{michels06prb}

The paper is organized as follows: To start with, we summarize in Sec.~\ref{satsec} well-known results for the magnetic SANS cross section of a fully saturated two-phase particle-matrix-type ferromagnet. Section~\ref{mumag} describes the micromagnetic model and provides approximate expressions for the Fourier coefficients of the magnetization, which are used in Sec.~\ref{sans} in order to compute the magnetic SANS cross section; here, we focus on the two most relevant scattering geometries where the externally applied magnetic field is perpendicular or parallel to the incident neutron beam. Section~\ref{corr} discusses the results in real space in terms of the corresponding spin-spin correlation function. Finally, Sec.~\ref{sum} summarizes the main findings of this work.

\section{Saturated state \label{satsec}}

It is our aim to derive an approximate expression for the magnetic microstructure and the ensuing magnetic SANS cross section of a two-phase nanocomposite ferromagnet, i.e., a bulk material which consists of a distribution of ferromagnetic nanoparticles that are embedded in and magnetically exchange-coupled to a ferromagnetic matrix; all particles are assumed to possess a saturation magnetization of $M_p$, whereas the saturation magnetization of the matrix phase is denoted with $M_m$. We remind the reader that for small-angle scattering the discrete atomic structure of both particle and matrix phase is generally of no relevance. Therefore, as far as magnetic SANS is concerned, the magnetization state may be represented by a continuous magnetization vector field $\mathbf{M}(\mathbf{r})$. Before discussing in detail in Sec.~\ref{mumag} the micromagnetic model of such a two-phase nanocomposite structure, we find it instructive to consider first the (somewhat idealized) fully saturated state.

If, for instance, the external magnetic field $\mathbf{H}_0 \parallel \mathrm{e}_z$ is large enough to fully align the magnetic moments of the sample, then the local magnetization simplifies to $\mathbf{M}(\mathbf{r}) = (0, 0, M_z)$, where the longitudinal component $M_z$ is explicitely dependent on the position $\mathbf{r}$ inside the material, i.e.,
\begin{equation}
\label{ms}
M_z(\mathbf{r}) = M_{p} \, u(\mathbf{r}) + M_{m} \, [1 - u(\mathbf{r})] \,\,,
\end{equation}
where $u(\mathbf{r}) = 1$ inside the particle phase and $u(\mathbf{r}) = 0$ inside the matrix phase. By assuming that the function $M_z(\mathbf{r})$ can be expressed as a Fourier integral,
\begin{equation}
\label{mzfourierdef1}
M_z(\mathbf{r}) = \frac{1}{(2\pi)^{3/2}} \int \widetilde{M}_z(\mathbf{q}) \, \exp(i \mathbf{q} \mathbf{r}) \, d^3q \,\,,
\end{equation}
\begin{equation}
\label{mzfourierdef2}
\widetilde{M}_z(\mathbf{q}) = \frac{1}{(2\pi)^{3/2}} \int M_z(\mathbf{r}) \, \exp(- i \mathbf{q} \mathbf{r}) \, d^3r \,\,,
\end{equation}
it is straightforward to obtain the following limiting cases: At $\mathbf{q} = 0$, one obtains the macroscopic saturation magnetization $M_s$ of the sample (which can be measured with a magnetometer),
\begin{eqnarray}
\label{mzaverage1}
M_s = \langle M_z \rangle = V^{-1} \int M_z(\mathbf{r}) \, d^3r \nonumber = \\
\frac{(2\pi)^{3/2}}{V} \widetilde{M}_z(\mathbf{q} = 0) = M_p \, p + M_m \, (1 - p) \,\,,
\end{eqnarray}
where $V$ is the sample volume and $p$ equals the volume fraction of the particles. At $\mathbf{q} \neq 0$, we obtain \cite{schloemann67}
\begin{equation}
\label{mzaverage2}
\widetilde{M}_z(\mathbf{q}) = \frac{\Delta M}{(2\pi)^{3/2}} \, I_p(\mathbf{q}) \,\,,
\end{equation}
where $\Delta M = M_p - M_m$ denotes the jump in the magnetization magnitude at the particle-matrix interface and the function $I_p(\mathbf{q})$ contains information about the size, the size distribution, the shape and the arrangement of the particles (compare Eq.~(\ref{mzsquareaverage}) below).

In the monodisperse and dilute limit we find for a single spherical particle (with radius $R$)
\begin{equation}
\label{mzaverage3}
\widetilde{M}_z(q) = \frac{\Delta M}{(2\pi)^{3/2}} \, 3 \, V_p \, \frac{j_1(q R)}{q R} \,\,,
\end{equation}
where $V_p = \frac{4\pi}{3} R^3$ and $j_1(q R)$ denotes the spherical Bessel function of first order. By using Eq.~(\ref{mzaverage3}) we can immediately write down (in the monodisperse and dilute limit) the magnetic SANS cross section $d \Sigma_M /d \Omega$ for a collection of $N_p$ saturated particles in a saturated matrix. For the particular scattering geometry where the wave vector $\mathbf{k}_0$ of the incident neutron beam is perpendicular to the applied magnetic field $\mathbf{H}_0$, we obtain the well-known expression
\begin{equation}
\label{sigmasansperpsat}
\frac{d \Sigma_M}{d \Omega}(\mathbf{q}) = \frac{N_p}{V} \, \Delta\rho^2_{\mathrm{mag}} \, V_p^2 \, F^2(q R) \, \sin^2\theta \,\,,
\end{equation}
where $\Delta\rho^2_{\mathrm{mag}} = b^2_H (\Delta M)^2$ represents the magnetic scattering-length density contrast, $F(q R) = 3 \, \frac{j_1(q R)}{q R}$ is the form factor of the sphere, and $\theta$ denotes the angle between the scattering vector $\mathbf{q}$ and $\mathbf{H}_0$ (compare Eq.~(\ref{sigmasansperp}) below).

Deviations from the saturated state result in the emergence of transversal magnetization components. The associated so-called spin-misalignment scattering will be addressed in the following sections.

\section{Micromagnetic model \label{mumag}}

Analytical micromagnetic calculations of the type presented here have already been carried out by other authors (e.g., Refs.~\onlinecite{schloemann67,schloemann71,kron77}). We employ the micromagnetic approach for computing (in the high-field limit) the differential magnetic scattering cross section (see Sec.~\ref{sans}). In the following we summarize the basic micromagnetic equations.

We are interested in the elastic spin-misalignment scattering which results from the static magnetic microstructure of a two-phase nanocomposite sample. Therefore, we start our analysis by writing down Brown's balance-of-torques equation \cite{brown,aharonibook,kronfahn03}
\begin{equation}
\label{torque}
\mathbf{M}(\mathbf{r}) \times \mathbf{H}_{\mathrm{eff}}(\mathbf{r}) = 0 \,\,,
\end{equation}
which expresses the fact that at static equilibrium the torque on the magnetization vector field $\mathbf{M}(\mathbf{r})$ due to an effective magnetic field $\mathbf{H}_{\mathrm{eff}}(\mathbf{r})$ vanishes everywhere. In the micromagnetic model we assume a uniform exchange interaction but an explicitely wave-vector-dependent longitudinal magnetization (see below). The effective field
\begin{equation}
\label{heff}
\mathbf{H}_{\mathrm{eff}}(\mathbf{r}) = \mathbf{H}_0 + \mathbf{H}_d(\mathbf{r}) + \mathbf{H}_p(\mathbf{r}) + \frac{2 A}{\mu_0 M_s^2} \nabla^2 \mathbf{M}(\mathbf{r}) \,\,
\end{equation}
is composed of a uniform applied magnetic field $\mathbf{H}_0$, of the magnetostatic field $\mathbf{H}_d(\mathbf{r})$, of the magnetic anisotropy field $\mathbf{H}_p(\mathbf{r})$, and of the exchange field [last term on the right hand side of Eq.~(\ref{heff})]; $\mu_0 = 4\pi \times 10^{-7} \, \mathrm{T m / A}$ is the permeability of free space, $\nabla = \mathbf{e}_x \, \partial / \partial x  + \mathbf{e}_y \, \partial / \partial y + \mathbf{e}_z \, \partial / \partial z$, where $\mathbf{e}_x$, $\mathbf{e}_y$, and $\mathbf{e}_z$ represent the unit vectors along the Cartesian laboratory axes. The parameter $A$ denotes the exchange-stiffness constant.

In the following we assume the material to be nearly saturated along $\mathbf{H}_0 \parallel \mathbf{e}_z$, i.e., we write
\begin{equation}
\label{mrealdef}
\mathbf{M}(\mathbf{r}) = M_x(\mathbf{r}) \, \mathbf{e}_x + M_y(\mathbf{r}) \, \mathbf{e}_y + M_z(\mathbf{r}) \, \mathbf{e}_z \,\,
\end{equation}
with $M_x \ll M_z$ and $M_y \ll M_z$ (small-misalignment approximation). The local saturation magnetization is assumed to differ only slightly from its spatial average, i.e., $M_z(\mathbf{r}) \cong \langle M_z \rangle = M_s$. Note that the jump $\Delta M$ in the longitudinal magnetization enters the calculation in $\mathbf{q}$-space via the expression for the Fourier coefficient of the magnetostatic field (see below). Furthermore, we assume that the anisotropy-energy density $\omega = \omega(\mathbf{r}, \mathbf{M})$ depends only linearly on the components of the magnetization. \cite{brown} As a consequence, the resulting anisotropy field $\mathbf{H}_p = - \mu_0^{-1} (\partial \omega / \partial M_x, \partial \omega / \partial M_y, \partial \omega / \partial M_z)$ is independent of $\mathbf{M}$ and, therefore, also independent of the applied magnetic field, implying that near saturation $\mathbf{H}_p = \mathbf{H}_p(\mathbf{r})$. Due to the micromagnetic constraint $|\mathbf{M}| = M_s$, an anisotropy-energy density of the form $\omega = \omega(\mathbf{r}, M_x, M_y, M_z)$ may be re-expressed as $\omega = \omega(\mathbf{r}, M_x, M_y)$ with the consequence that only two independent components of $\mathbf{H}_p$ exist. In the approach-to-saturation regime, when $\mathbf{M}$ is nearly aligned parallel to the external magnetic field $\mathbf{H}_0$, only those components of $\mathbf{H}_p$ which are normal to $\mathbf{H}_0$ are physically effective in producing a torque on the magnetization.

Basic magnetostatics prescribes that $\nabla \cdot (\mathbf{H}_0 + \mathbf{H}_d) = - \nabla \cdot \mathbf{M}$ and that $\nabla \times (\mathbf{H}_0 + \mathbf{H}_d) = 0$. The magnetostatic field $\mathbf{H}_d(\mathbf{r})$ can be written as the sum of the surface demagnetizing field $\mathbf{H}_d^s$ and of the magnetostatic field $\mathbf{H}_d^b$ which is related to volume charges, i.e., $\mathbf{H}_d = \mathbf{H}_d^s + \mathbf{H}_d^b$. In the high-field limit and for samples with an ellipsoidal shape with $\mathbf{H}_0$ directed along a principal axis of the ellipsoid, one may approximate the demagnetizing field due to the surface charges by the uniform field $\mathbf{H}_d^s = - N M_s \mathbf{e}_z$, where $N$ denotes the corresponding demagnetizing factor. In Fourier space (at $\mathbf{q} \neq 0$) the above magnetostatic relations suggest the following expression for the Fourier coefficient $\mathbf{h}_d^b(\mathbf{q})$ of $\mathbf{H}_d^b(\mathbf{r})$, \cite{herring51}
\begin{equation}
\label{hdbdef}
\mathbf{h}_d^b(\mathbf{q}) = - \frac{\mathbf{q} \, [\mathbf{q} \, \mathbf{\widetilde{M}}(\mathbf{q})]}{q^2} \,\,,
\end{equation}
\begin{equation}
\label{hdbfourierdef}
\mathbf{H}_d^b(\mathbf{r}) = \frac{1}{(2\pi)^{3/2}} \int \mathbf{h}_d^b(\mathbf{q}) \, \exp(i \mathbf{q} \mathbf{r}) \, d^3q \,\,.
\end{equation}
$\mathbf{\widetilde{M}}(\mathbf{q})$ represents the Fourier transform of the magnetization $\mathbf{M}(\mathbf{r})$,
\begin{equation}
\label{mfourierdef}
\mathbf{M}(\mathbf{r}) = \frac{1}{(2\pi)^{3/2}} \int \mathbf{\widetilde{M}}(\mathbf{q}) \, \exp(i \mathbf{q} \mathbf{r}) \, d^3q \,\,.
\end{equation}

Likewise, the Fourier transform $\mathbf{h}(\mathbf{q}) = (h_x(\mathbf{q}), h_y(\mathbf{q}), 0)$ of the magnetic anisotropy field $\mathbf{H}_p(\mathbf{r})$ is introduced as
\begin{equation}
\label{Hpfourierdef}
\mathbf{H}_p(\mathbf{r}) = \frac{1}{(2\pi)^{3/2}} \int \mathbf{h}(\mathbf{q}) \, \exp(i \mathbf{q} \mathbf{r}) \, d^3q \,\,.
\end{equation}
The details of the sample's microstructure (e.g., grain size, lattice strain, crystallographic texture) are included in $\mathbf{H}_p(\mathbf{r})$. \cite{jprb2001}

By inserting Eqs.~(\ref{mrealdef})$-$(\ref{Hpfourierdef}) into Eqs.~(\ref{torque}) and (\ref{heff}), we obtain for a general orientation of the wave vector $\mathbf{q} = (q_x, q_y, q_z)$ the following expressions for $\widetilde{M}_x(\mathbf{q})$ and $\widetilde{M}_y(\mathbf{q})$:
\begin{widetext}
\begin{eqnarray}
\label{solmx}
\widetilde{M}_x(\mathbf{q}) = M_s \, \frac{ \left( h_x - \widetilde{M}_z \frac{q_x q_z}{q^2} \right) \left( H_{\mathrm{eff}} + M_s \frac{q_y^2}{q^2} \right) - M_s \frac{q_x q_y}{q^2} \left( h_y - \widetilde{M}_z \frac{q_y q_z}{q^2} \right)}{H_{\mathrm{eff}} \left( H_{\mathrm{eff}} + M_s \frac{q_x^2 + q_y^2}{q^2} \right)} \,\,, \\
\label{solmy}
\widetilde{M}_y(\mathbf{q}) = M_s \, \frac{\left( h_y - \widetilde{M}_z \frac{q_y q_z}{q^2} \right) \left( H_{\mathrm{eff}} + M_s \frac{q_x^2}{q^2} \right) - M_s \frac{q_x q_y}{q^2} \left( h_x - \widetilde{M}_z \frac{q_x q_z}{q^2} \right)}{H_{\mathrm{eff}} \left( H_{\mathrm{eff}} + M_s \frac{q_x^2 + q_y^2}{q^2} \right)} \,\,.
\end{eqnarray}
\end{widetext}

The terms in Eqs.~(\ref{solmx}) and (\ref{solmy}) which contain the Fourier coefficient $\widetilde{M}_z(\mathbf{q}) \propto \Delta M$ model the influence of the two-phase magnetic microstructure on the magnetic SANS and are not contained in the corresponding expressions for the single-phase case (compare Eqs.~(2.15) in Ref.~\onlinecite{michels08rop}). The quantity
\begin{equation}
\label{heffdef}
H_{\mathrm{eff}}(q, H_i) = H_i \left( 1 + l_H^2 q^2 \right)
\end{equation}
is the effective magnetic field, which depends on the internal field $H_i = H_0 - N M_s$, on $q = |\mathbf{q}|$, and on the exchange length of the field $l_H(H_i) = \sqrt{2 A / (\mu_0 M_s H_i)}$; the length scale $l_H$ characterizes the range over which perturbations in $\mathbf{M}$ decay (see Sec.~\ref{corr} below). \cite{michels03prl,michelsprb2010}

In the derivation of Eqs.~(\ref{solmx}) and (\ref{solmy}) terms of higher than linear order in $M_x$ or $M_y$ have been neglected, including terms such as $H_{d,i}^b \, M_x$ or $H_{d,i}^b \, M_y$ where $i \in \{x, y, z\}$. \cite{aharonibook} Besides this small-misalignment approximation, we have introduced two further approximations: (i) The exchange interaction is assumed to be homogeneous, i.e., jumps in $A$ between the two magnetic phases have not been taken into account. Such an approximation is permissable as long as exchange fluctuations are not too large, in particular, for soft magnetic materials. \cite{skomski2003} (ii) The function $\widetilde{M}_z(\mathbf{q})$, which models the influence of the two-phase magnetic microstructure on the magnetic SANS, is introduced into our theory only in $\mathbf{q}$-space via $\mathbf{h}_d^b(\mathbf{q}) = - \mathbf{q} \, [\mathbf{q} \, \mathbf{\widetilde{M}}(\mathbf{q})] / q^2$ [Eq.~(\ref{hdbdef})]. This is an approximation, since in real space we have assumed that $M_z \cong M_s =$~constant, and hence $\widetilde{M}_z(\mathbf{q}) \propto \delta(\mathbf{q})$ would result, as is appropriate for a homogeneous single-phase material. However, by explicitly considering the $\mathbf{q} \neq 0$ Fourier coefficients of $\widetilde{M}_z$ [Eq.~(\ref{mzaverage2})], it becomes possible to straightforwardly include the jump in the longitudinal magnetization at the particle-matrix interface, and one avoids the otherwise necessary calculation of convolution products. \cite{schloemann67,schloemann71,kron77}

For the following discussion of magnetic SANS it is of interest to consider special projections of Eqs.~(\ref{solmx}) and (\ref{solmy}) into the plane of the 2D detector. The two scattering geometries which are of particular relevance to experiment have the external magnetic field $\mathbf{H}_0$ either perpendicular or parallel to the wave vector $\mathbf{k}_0$ of the incoming neutron beam. For $\mathbf{k}_0 \parallel \mathbf{e}_x$ and $\mathbf{H}_0 \parallel \mathbf{e}_z$, the scattering vector can be approximated as $\mathbf{q} \cong (0, q_y, q_z)$, i.e., $q_x \cong 0$, and Eqs.~(\ref{solmx}) and (\ref{solmy}) reduce to
\begin{eqnarray}
\label{solmxqx0}
\widetilde{M}_x(\mathbf{q}) &=& M_s \, \frac{h_x(\mathbf{q})}{H_{\mathrm{eff}}} \,\,, \\
\label{solmyqx0}
\widetilde{M}_y(\mathbf{q}) &=& M_s \, \frac{h_y(\mathbf{q}) - \widetilde{M}_z(\mathbf{q}) \frac{q_y q_z}{q^2}}{H_{\mathrm{eff}} + M_s \frac{q^2_y}{q^2}} \,\,.
\end{eqnarray}

For $\mathbf{k}_0 \parallel \mathbf{H}_0 \parallel \mathbf{e}_z$, $\mathbf{q} \cong (q_x, q_y, 0)$, i.e., $q_z \cong 0$, and the results for the Fourier coefficients simplify to
\begin{eqnarray}
\label{solmxqz0}
\widetilde{M}_x(\mathbf{q}) &=& M_s \, \frac{h_x \, \left( H_{\mathrm{eff}} + M_s \frac{q^2_y}{q^2} \right) - h_y \, M_s \frac{q_x q_y}{q^2}}{H_{\mathrm{eff}} \, \left( H_{\mathrm{eff}} + M_s \frac{q^2_x + q^2_y}{q^2} \right)} \,\,, \\
\label{solmyqz0}
\widetilde{M}_y(\mathbf{q}) &=& M_s \, \frac{h_y \, \left( H_{\mathrm{eff}} + M_s \frac{q^2_x}{q^2} \right) - h_x \, M_s \frac{q_x q_y}{q^2}}{H_{\mathrm{eff}} \, \left( H_{\mathrm{eff}} + M_s \frac{q^2_x + q^2_y}{q^2} \right)} \,\,.
\end{eqnarray}
Equations~(\ref{solmxqz0}) and (\ref{solmyqz0}) are similar to Eqs.~(2.15) in Ref.~\onlinecite{michels08rop}.

The high-field solutions for $\widetilde{M}_x(\mathbf{q})$ and $\widetilde{M}_y(\mathbf{q})$ can be seen as a sum of products of components of the anisotropy-field Fourier coefficient $\mathbf{h}(\mathbf{q})$ and $\widetilde{M}_z(\mathbf{q})$ with micromagnetic functions which contain the effective magnetic field $H_{\mathrm{eff}}$ and terms that depend on the orientation of the wavevector (e.g., $M_s \, q_y^2 / q^2$). The convolution theorem then implies that the magnetic microstructure in real space, $\mathbf{M}(\mathbf{r})$, is tantamount to a complicated convolution product between the corresponding real-space functions. As a consequence, sharp features in the nuclear or anisotropy-field microstructure are washed out and smoothly-varying magnetization profiles are at the origin of the related spin-misalignment scattering (compare, e.g., Figs.~$2-4$ in Ref.~\onlinecite{michelsprb2010}). Consistent with this notion is the observation of power-law exponents significantly larger than 4 (see Fig.~\ref{fig4} below) and the finding that the slope of the correlation function at the origin vanishes (see Fig.~\ref{fig5} below). \cite{porod}

\section{SANS cross section for unpolarized neutrons \label{sans}}

Although the following derivation of the magnetic SANS cross section is aimed to be self-contained, we refer the reader to Refs.~\onlinecite{jnist,jprb2001} for further details.

\subsection{$\mathbf{k}_0 \perp \mathbf{H}_0$}

For the transversal scattering geometry, one can express the total nuclear and magnetic SANS cross section $d \Sigma / d \Omega$ at scattering vector $\mathbf{q}$ as \cite{michels08rop}
\begin{widetext}
\begin{eqnarray}
\label{sigmasansperp}
\frac{d \Sigma}{d \Omega}(\mathbf{q}) = \frac{8 \pi^3}{V} \, b_H^2 \left( \frac{|\widetilde{N}|^2}{b_H^2} + |\widetilde{M}_x|^2 + |\widetilde{M}_y|^2 \cos^2\theta + |\widetilde{M}_z|^2 \sin^2\theta - (\widetilde{M}_y \widetilde{M}_z^{\ast} + \widetilde{M}_y^{\ast} \widetilde{M}_z) \sin\theta \cos\theta \right) \,\,.
\end{eqnarray}
\end{widetext}
Here, $V$ is the scattering volume, $b_H = 2.9 \times 10^{8} \, \mathrm{A^{-1} m^{-1}}$, $\widetilde{N}(\mathbf{q})$ is the nuclear scattering amplitude, and $\theta$ denotes the angle between $\mathbf{H}_0$ and $\mathbf{q} \cong q \, (0, \sin\theta, \cos\theta)$; $c^{*}$ is a quantity complex-conjugated to $c$. The magnetic form factor in the expression for the magnetic scattering amplitude ($\propto b_H$) was set to unity, which is permissible along the forward direction.

By inserting Eqs.~(\ref{solmxqx0}) and (\ref{solmyqx0}) into Eq.~(\ref{sigmasansperp}) one can express $d \Sigma / d \Omega$ as
\begin{equation}
\label{sigmasansperp2d}
\frac{d \Sigma}{d \Omega}(\mathbf{q}) = \frac{d \Sigma_{\mathrm{res}}}{d \Omega}(\mathbf{q}) + \frac{d \Sigma_M}{d \Omega}(\mathbf{q}) \,\,,
\end{equation}
where
\begin{equation}
\label{sigmaresperp}
\frac{d \Sigma_{\mathrm{res}}}{d \Omega}(\mathbf{q}) = \frac{8 \pi^3}{V} \, b_H^2 \left( \frac{|\widetilde{N}|^2}{b_H^2} + |\widetilde{M}_z|^2 \sin^2\theta \right)
\end{equation}
represents the (nuclear and magnetic) residual SANS cross section, which may be measured at complete magnetic saturation. Subtraction of $d \Sigma_{\mathrm{res}} / d \Omega$ from a measurement of $d \Sigma / d \Omega$ at a lower field yields the spin-misalignment SANS cross section $d \Sigma_M / d \Omega$, which contains the scattering contributions due to the misaligned spins,
\begin{equation}
\label{sigmasmperp}
\frac{d \Sigma_M}{d \Omega}(\mathbf{q}) = S_H(\mathbf{q}) \, R_H(q, \theta, H_i) + S_M(\mathbf{q}) \, R_M(q, \theta, H_i) \,\,.
\end{equation}

Following Weissm\"uller {\it et al.} (Ref.~\onlinecite{jnist}), we have here introduced the scattering function of the anisotropy field,
\begin{equation}
\label{shdef}
S_H(\mathbf{q}) = \frac{8 \pi^3}{V} \, b_H^2 \, h^2(\mathbf{q}) \,\,,
\end{equation}
the scattering function of the longitudinal magnetization,
\begin{equation}
\label{smdef}
S_M(\mathbf{q}) = \frac{8 \pi^3}{V} \, b_H^2 \, \widetilde{M}_z^2(\mathbf{q}) \,\,,
\end{equation}
and the corresponding (dimensionless) micromagnetic response functions,
\begin{equation}
\label{rhdefperp}
R_H(q, \theta, H_i) = \frac{p^2}{2} \left( 1 + \frac{\cos^2\theta}{\left( 1 + p \sin^2\theta \right)^2} \right) \,\,,
\end{equation}
\begin{equation}
\label{rmdefperp}
R_M(q, \theta, H_i) = \frac{p^2 \, \sin^2\theta \cos^4\theta}{\left( 1 + p \sin^2\theta \right)^2} + \frac{2 p \, \sin^2\theta \cos^2\theta}{1 + p \sin^2\theta} \,\,,
\end{equation}
where $p(q, H_i) = M_s/H_{\mathrm{eff}}(q, H_i)$. We note that the term $S_M \times R_M$ in Eq.~(\ref{sigmasmperp}) [and Eq.~(\ref{sigmasmperpradav})] is not contained in the expression for the single-phase material case (homogeneous saturation magnetization), where \cite{michels08rop}
\begin{equation}
\label{singlephaseperp}
\frac{d \Sigma_M}{d \Omega}(\mathbf{q}) = S_H(\mathbf{q}) \, R_H(q, \theta, H_i) \,\,.
\end{equation}

Near magnetic saturation, $S_H$ and $S_M$ are both approximately independent of the applied magnetic field and contain, respectively, information on the strength and on the spatial structure of the magnetic anisotropy field \cite{jprb2001} as well as on the magnitude of the jump of the magnetization at the particle-matrix interface. The geometry of the microstructure is contained in $S_H$ and $S_M$. Both response functions depend on the magnitude and orientation of the scattering vector, on the applied field, and on the magnetic interaction parameters.

As mentioned previously, the effects of crystallographic texture or of other forms of anisotropy on $d \Sigma_M / d \Omega$ enter the theory mainly through the magnetic anisotropy field $\mathbf{H}_p(\mathbf{r})$. In deriving Eqs.~(\ref{sigmasansperp2d})$-$(\ref{rmdefperp}) we have assumed that the corresponding anisotropy-field Fourier coefficient $\mathbf{h}(\mathbf{q}) = (h(\mathbf{q}) \cos\beta, h(\mathbf{q}) \sin\beta, 0)$ is isotropically distributed in the plane perpendicular to $\mathbf{H}_0$, in other words, the vector $\mathbf{h}(\mathbf{q})$ takes on all orientations (i.e., angles $\beta$) with equal probability. This assumption allows us to average the response functions over $\beta$, i.e., $1/(2\pi) \int_0^{2\pi} (...) \, d\beta$, which results in Eqs.~(\ref{rhdefperp}) and (\ref{rmdefperp}); note that interference terms $\propto h_y \, \widetilde{M}_z = h \sin\beta \, \widetilde{M}_z$ vanish in this averaging procedure (see also Sec.~\ref{average} below). The case of a texture in the orientation of the anisotropy field is treated in the Appendix of Ref.~\onlinecite{jprb2001}.

Furthermore, by assuming that both Fourier coefficients $h^2$ and $\widetilde{M}_z^2$ depend only on the magnitude $q$ of the scattering vector $\mathbf{q}$, and not on its orientation $\theta$, one may average Eq.~(\ref{sigmasmperp}) [and Eq.~(\ref{singlephaseperp})] with respect to the angle $\theta$, i.e., $1/(2\pi) \int_0^{2\pi} (...) \, d\theta$, which results in
\begin{eqnarray}
\label{sigmasmperpradav}
\frac{d \Sigma_M}{d \Omega}(q) = S_H(q) \, R_H(q, H_i) + S_M(q) \, R_M(q, H_i) \,\,,
\end{eqnarray}
where
\begin{equation}
\label{rhdefperpradav}
R_H(q, H_i) = \frac{p^2}{4} \left( 2 + \frac{1}{\sqrt{1 + p}} \right)\,\,,
\end{equation}
\begin{equation}
\label{rmdefperpradav}
R_M(q, H_i) = \frac{\sqrt{1 + p} - 1}{2} \,\,.
\end{equation}
Equation~(\ref{sigmasmperpradav}) may be compared to experimental data for $d \Sigma_M / d \Omega$ in order to obtain the functions $S_H$ and $S_M$ and a value for the average exchange-stiffness constant $A$.

\subsection{$\mathbf{k}_0 \parallel \mathbf{H}_0$}

For the longitudinal SANS geometry, $d \Sigma / d \Omega$ reads
\begin{widetext}
\begin{eqnarray}
\label{sigmasanspara}
\frac{d \Sigma}{d \Omega}(\mathbf{q}) = \frac{8 \pi^3}{V} \, b_H^2 \left( \frac{|\widetilde{N}|^2}{b_H^2} + |\widetilde{M}_x|^2 \sin^2\theta + |\widetilde{M}_y|^2 \cos^2\theta + |\widetilde{M}_z|^2 - (\widetilde{M}_x \widetilde{M}_y^{\ast} + \widetilde{M}_x^{\ast} \widetilde{M}_y) \sin\theta \cos\theta \right) \,\,,
\end{eqnarray}
\end{widetext}
where $\mathbf{q} \cong q \, (\cos\theta, \sin\theta, 0)$ and $\theta = \angle(\mathbf{e}_x, \mathbf{q})$. Inserting Eqs.~(\ref{solmxqz0}) and (\ref{solmyqz0}) into Eq.~(\ref{sigmasanspara}) and averaging over the orientations of the magnetic anisotropy field (angle $\beta$) results in
\begin{eqnarray}
\label{sigmasanspara2d}
\frac{d \Sigma}{d \Omega}(\mathbf{q}) = \frac{d \Sigma_{\mathrm{res}}}{d \Omega}(\mathbf{q}) + \frac{d \Sigma_M}{d \Omega}(\mathbf{q}) \,\,,
\end{eqnarray}
where the residual SANS cross section equals
\begin{equation}
\label{sigmarespara}
\frac{d \Sigma_{\mathrm{res}}}{d \Omega}(\mathbf{q}) = \frac{8 \pi^3}{V} \, b_H^2 \left( \frac{|\widetilde{N}|^2}{b_H^2} + |\widetilde{M}_z|^2 \right) \,\,,
\end{equation}
and the spin-misalignment SANS cross section is expressed as
\begin{eqnarray}
\label{sigmasmpara}
\frac{d \Sigma_M}{d \Omega}(\mathbf{q}) = S_H(\mathbf{q}) \, R_H(q, H_i) \,\,,
\end{eqnarray}
with an isotropic (i.e., $\theta$-independent) response function,
\begin{equation}
\label{rdefpararadav}
R_H(q, H_i) = \frac{p^2}{2} \,\,.
\end{equation}
We note that in the longitudinal SANS geometry $d \Sigma_M / d \Omega$ is independent of $\widetilde{M}_z$ and equals the expression for the single-phase case [compare Eq.~(33) in Ref.~\onlinecite{jnist}]. In other words, the two-phase nature of the underlying microstructure is (for $\mathbf{k}_0 \parallel \mathbf{H}_0$) only contained in the residual SANS, and not in $d \Sigma_M / d \Omega$.

\subsection{Comment on the average of $d \Sigma_M / d \Omega$ over many statistically uncorrelated defects \label{average}}

The sample volume which is probed by the neutrons typically contains many defects (e.g., particles), each one having a different orientation and/or magnitude of the magnetic anisotropy field. For ferromagnets with a nonuniform saturation magnetization, each particle or crystallite ``$j$'' is additionally characterized by the form-factor function $\widetilde{M}_{z,j}$.

In order to discuss the statistical average over the spin-misalignment SANS cross sections of many defects, we ignore for the moment the $\Delta M$ fluctuations, i.e., we consider the case of a \emph{homogeneous single-phase ferromagnet} with uniform values of $A$ and $M_s$. Here, the dominating source of spin disorder (in the approach-to-saturation regime) is related to spatially inhomogeneous magnetic anisotropy fields. We follow the arguments of Weissm\"uller \cite{jnist,jprb2001} and assume that the total anisotropy-field Fourier coefficient of the sample, $\mathbf{h}(\mathbf{q})$, can be expressed as the sum of the anisotropy-field amplitudes of the individual defects,
\begin{equation}
\label{hjsuperpo}
\mathbf{h}(\mathbf{q}) = \sum_j \mathbf{h}_j(\mathbf{q}) \,\,.
\end{equation}
If the $\mathbf{h}_j$ of the individual defects are statistically uncorrelated (random anisotropy), then the expectation value of $|\mathbf{h}(\mathbf{q})|^2$ reduces to the sum over the individual $|\mathbf{h}_j|^2$,
\begin{equation}
\label{hjsuperposquare}
|\mathbf{h}(\mathbf{q})|^2 = \sum_j |\mathbf{h}_j(\mathbf{q})|^2 \,\,.
\end{equation}
Since---within the linear micromagnetic approximation---$d \Sigma_M / d \Omega$ is proportional to $|\mathbf{h}(\mathbf{q})|^2$ [compare Eq.~\ref{singlephaseperp}], it immediately follows that the above additivity also transfers to the total spin-misalignment SANS cross section of the sample. In other words,
\begin{equation}
\label{sigmasuperpo}
\frac{d \Sigma_M}{d \Omega} = \sum_j \frac{d \Sigma_{M,j}}{d \Omega} = \sum_j S_{H,j} \, R_{H,j} \,\,,
\end{equation}
when homogeneous single-phase ferromagnets are considered. \cite{jnist,jprb2001} This is in contrast to nuclear SANS or SAXS, where the decomposition of the overall cross section into the sum over cross sections of individual particles is only permissible for small particle volume fractions.

Including now the $\Delta M$ fluctuations into the discussion, inspection of Eq.~(\ref{solmyqx0}) shows that interference terms $\propto h_{y,j} \, \widetilde{M}_{z,j}$ appear in the spin-misalignment SANS cross section [compare Eq.~(\ref{sigmasansperp})]. However, in deriving the final expression for $d \Sigma_M / d \Omega$ [Eq.~(\ref{sigmasmperp})] these terms cancel when the averaging procedure over the (isotropic) orientations of the anisotropy field (angle $\beta$) are carried out, with the result that the scattering contributions due to inhomogeneous anisotropy fields $S_H \times R_H$ and magnetostatic fluctuations $S_M \times R_M$ are additive. Taking into account interparticle interference effects, the general expression for $|\widetilde{M}_z|^2$ (at $\mathbf{q} \neq 0$) may be expressed as \cite{schloemann67}
\begin{equation}
\label{mzsquareaverage}
|\widetilde{M}_z(\mathbf{q})|^2 = \frac{(\Delta M)^2}{8\pi^{3}} \, \left| \sum_{j=1}^{N_p} V_{p,j} \, F_j(\mathbf{q}) \, \exp(- i \mathbf{q} \, \mathbf{r}_j) \right|^2 \,\,,
\end{equation}
where $N_p$, $V_{p,j}$, $F_j$ and $\mathbf{r}_j$ denote, respectively, the number of particles, the volume, the form factor and the position vector of particle $j$ [compare Eq.~(\ref{mzaverage3})]. Equation~(\ref{mzsquareaverage}) can be employed in Eqs.~(\ref{sigmasmperp}) or (\ref{sigmasmperpradav}) in order to analyze experimental data.

\subsection{Graphical representation of $d \Sigma_M / d \Omega$}

\begin{figure}[tb]
\centering
\resizebox{0.65\columnwidth}{!}{\includegraphics{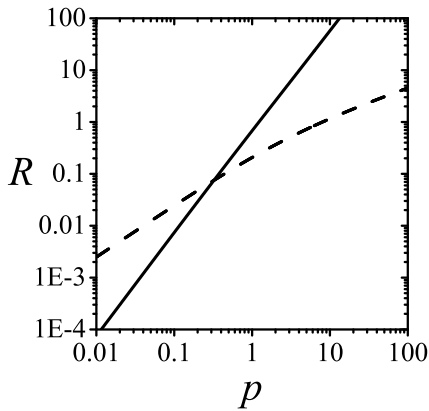}}
\caption{Micromagnetic response functions $R_H$ (Eq.~(\ref{rhdefperpradav}); solid line) and $R_M$ (Eq.~(\ref{rmdefperpradav}); dashed line) versus $p(q, H_i) = M_s/H_{\mathrm{eff}}(q, H_i)$ (log-log scale).}
\label{fig1}
\end{figure}

\begin{figure*}[tb]
\centering
\resizebox{1.75\columnwidth}{!}{\includegraphics{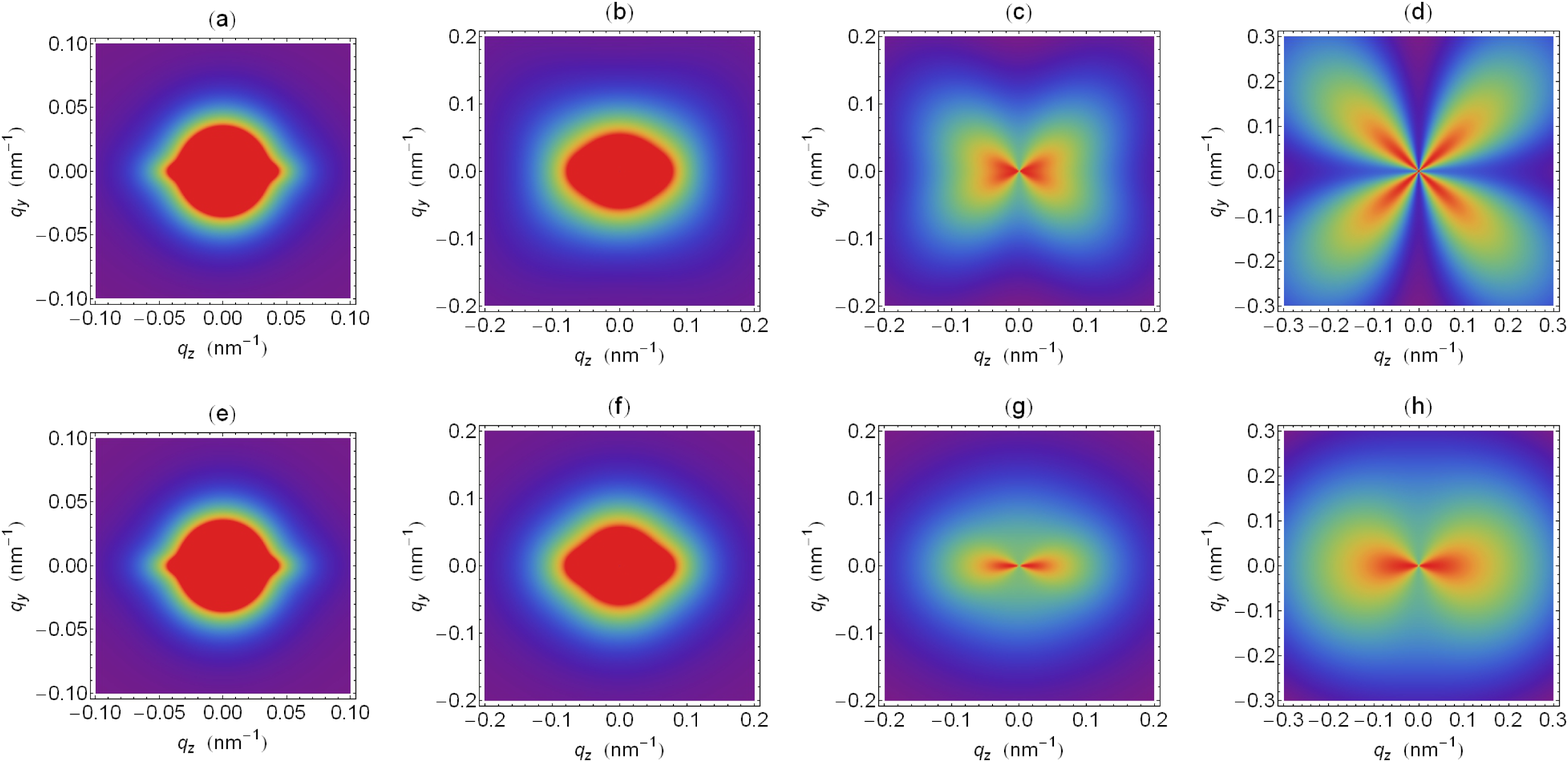}}
\caption{(Color online) (upper row) Contour plots of $d \Sigma_M / d \Omega$ (in arbitrary units) for $\mathbf{k}_0 \perp \mathbf{H}_0$ [Eq.~(\ref{sigmasmperp})]. $\mathbf{H}_0$ is horizontal. $H_p / \Delta M = 1$ ($S_H = S_M$). $H_i$ values (in T) from (a) to (d): 0.01; 0.2; 1.0; 10.0. (lower row) Field dependence of $d \Sigma_M / d \Omega$ for $\mathbf{k}_0 \perp \mathbf{H}_0$ for a homogeneous single-phase ferromagnet [Eq.~(\ref{singlephaseperp})], where $\Delta M = 0$. Field values and materials parameters in (e) to (h) are the same as in (a) to (d). Red color corresponds to ``high intensity'' and blue color to ``low intensity''.}
\label{fig2}
\end{figure*}

In order to graphically display $d \Sigma_M / d \Omega$ [Eq.~(\ref{sigmasmperp})] it is necessary to specify particular models for the Fourier coefficients $h(q)$ and $\widetilde{M}_z(q)$. For $\widetilde{M}_z(q)$, we use for simplicity the form factor of the sphere [Eq.~(\ref{mzaverage3})], which implies the neglect of interparticle interactions (dilute limit). We note, however, that the inclusion of the structure factor of the material, of other particle shapes (form factors), or of the particle-size distribution is straightforward [compare Eq.~(\ref{mzsquareaverage})]. \cite{pedersen97} For $h(q)$, we employ the model introduced by Weissm\"uller {\it et al.} (Eq.~(43) in Ref.~\onlinecite{jnist}), which considers a nanocrystalline material composed of spherical particles with a constant magnitude but random orientation of $\mathbf{H}_p$. The resulting expression for $h(q)$ is the form factor of the sphere, Eq.~(\ref{mzaverage3}), where $\Delta M$ is replaced by the magnitude $H_p$ of the magnetic anisotropy field. For the particle radius, we assume a value of $R = 5 \, \mathrm{nm}$ for both $\widetilde{M}_z(qR)$ and $h(qR)$. Since, then, $\widetilde{M}_z(qR)$ and $h(qR)$ and, hence, $S_M$ and $S_H$ are equal except for the prefactors, the ratio $H_p / \Delta M$ determines the properties of $d \Sigma_M / d \Omega$. Unless otherwise stated, we have used the following materials parameters: $A = 2.5 \times 10^{-11} \, \mathrm{J/m}$; $\mu_0 M_s = 1.5 \, \mathrm{T}$; $\mu_0 \Delta M = 0.25 \, \mathrm{T}$.

The spin-misalignment SANS cross section for $\mathbf{k}_0 \perp \mathbf{H}_0$ [Eq.~(\ref{sigmasmperp})] contains scattering contributions due to the magnetic anisotropy field, $S_H \times R_H$, and due to jumps in the longitudinal magnetization, $S_M \times R_M$. In Fig.~\ref{fig1} we plot both response functions $R_H$ [Eq.~(\ref{rhdefperpradav})] and $R_M$ [Eq.~(\ref{rmdefperpradav})] as a function of the dimensionless parameter $p$. Assuming that $S_H$ and $S_M$ are of comparable magnitude, it is seen that at large applied fields or large momentum transfers (when $p \ll 1$), $d \Sigma_M / d \Omega$ is dominated by the term $S_M \times R_M$, whereas at small fields and small momentum transfers (when $p \gg 1$), $d \Sigma_M / d \Omega$ is governed by the $S_H \times R_H$ contribution (see Fig.~\ref{fig2} below).

Figure~\ref{fig2} qualitatively displays the applied-field dependence of $d \Sigma_M / d \Omega$ for $\mathbf{k}_0 \perp \mathbf{H}_0$ [Eq.~(\ref{sigmasmperp})] and for a ratio of $H_p / \Delta M = 1$ ($S_H = S_M$); the results are compared with the $d \Sigma_M / d \Omega$ of a single-phase material with a uniform saturation magnetization [Eq.~(\ref{singlephaseperp})]. The $d \Sigma_M / d \Omega$ for the two-phase case [Figs.~\ref{fig2}(a)$-$(d)] reveal a strongly field-dependent angular anisotropy. At the largest fields, the pattern exhibits maxima roughly along the diagonals of the detector---the so-called ``clover-leaf'' anisotropy---previously observed in the Fe-based two-phase alloy NANO\-PERM (compare, e.g., Fig.~3 in Ref.~\onlinecite{michels06prb}). The position of the maxima in $d \Sigma_M / d \Omega$ depend on $q$ and $H_i$ (see also Fig.~11 in Ref.~\onlinecite{michels2013jmmm}). Such a type of clover-leaf anisotropy cannot be reproduced by the $d \Sigma_M / d \Omega$ for the single-phase case [Figs.~\ref{fig2}(e)$-$(h)]. Here, at large $q$ and $H_i$, one observes an elongation of the spin-misalignment scattering along the field direction [due to the $\cos^2\theta$ term in Eq.~(\ref{rhdefperp})], with a ``flying-saucer-type'' pattern taking over at small $q$ and $H_i$ [due to the $\sin^2\theta$ term in the denominator of Eq.~(\ref{rhdefperp})]. The sharp spike in Figs.~\ref{fig2}(a) and (e) is due to the magnetostatic interaction and was first predicted by Weissm\"uller {\it et al.} \cite{jnist}

\begin{figure}[tb]
\centering
\resizebox{1.0\columnwidth}{!}{\includegraphics{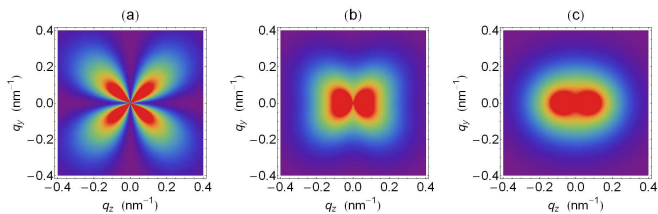}}
\caption{(Color online) Crossover from magnetostatic to anisotropy-field dominated scattering. Contour plots of $d \Sigma_M / d \Omega$ (in arbitrary units) for $\mathbf{k}_0 \perp \mathbf{H}_0$ [Eq.~(\ref{sigmasmperp})] at a fixed internal magnetic field of $\mu_0 H_i = 2.0 \, \mathrm{T}$. $\mathbf{H}_0$ is horizontal. Values of $H_p / \Delta M$ from (a) to (c): 0.2; 1.6; 8. Red color corresponds to ``high intensity'' and blue color to ``low intensity''.}
\label{fig3}
\end{figure}

Figure~\ref{fig3} shows $d \Sigma_M / d \Omega$ for $\mathbf{k}_0 \perp \mathbf{H}_0$ [Eq.~(\ref{sigmasmperp})] at a fixed internal field of $2.0 \, \mathrm{T}$ but for different magnitudes of the magnetic anisotropy field $H_p$ relative to the jump $\Delta M$ in the magnetization magnitude at the particle-matrix interface. One can clearly observe a transition from dipole-field dominated scattering, with a characteristic clover-leaf-type pattern [Fig.~\ref{fig3}(a)], to a more anisotropy-field dominated $\cos^2\theta$-type angular anisotropy of $d \Sigma_M / d \Omega$ [Fig.~\ref{fig3}(c)].

\begin{figure}[tb]
\centering
\resizebox{0.65\columnwidth}{!}{\includegraphics{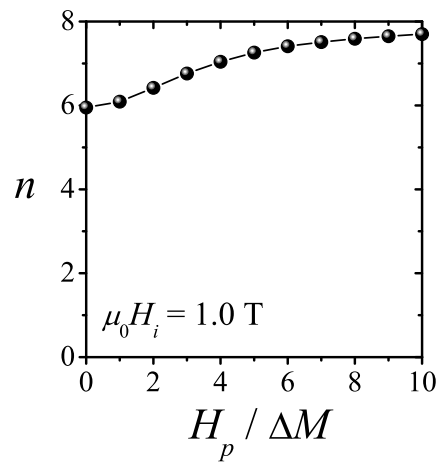}}
\caption{Variation of the power-law exponent $n$ in $d \Sigma_M / d \Omega = K/q^n$ with the ratio $H_p / \Delta M$ ($\mu_0 H_i = 1.0 \, \mathrm{T}$) ($\mathbf{k}_0 \perp \mathbf{H}_0$). The fits of the above function to the simulated data [Eq.~(\ref{sigmasmperpradav})] have been restricted to the interval $1.0 \, \mathrm{nm}^{-1} < q < 2.0 \, \mathrm{nm}^{-1}$. Line is guide to the eye.}
\label{fig4}
\end{figure}

Regarding the asymptotic power-law dependence of $d \Sigma_M / d \Omega$: For particles with sharp interfaces, both $h^2(q)$ and $\widetilde{M}^2_z(q)$ vary asymptotically as $q^{-4}$, as does the function $H^{-2}_{\mathrm{eff}}$ [compare Eq.~(\ref{heffdef})]. Taking these dependencies into account, it is readily verified that the anisotropy-field contribution to $d \Sigma_M / d \Omega$ varies as $S_H \times R_H \propto q^{-8}$, whereas $S_M \times R_M \propto q^{-6}$. Therefore, depending on the relative magnitude of both contributions to $d \Sigma_M / d \Omega$, one observes different asymptotic power-law exponents (see Fig.~\ref{fig4}). We note that other models for the anisotropy-field microstructure may result in different exponents.

\section{Correlation function of the spin misalignment \label{corr}}

\begin{figure}[tb]
\centering
\resizebox{0.85\columnwidth}{!}{\includegraphics{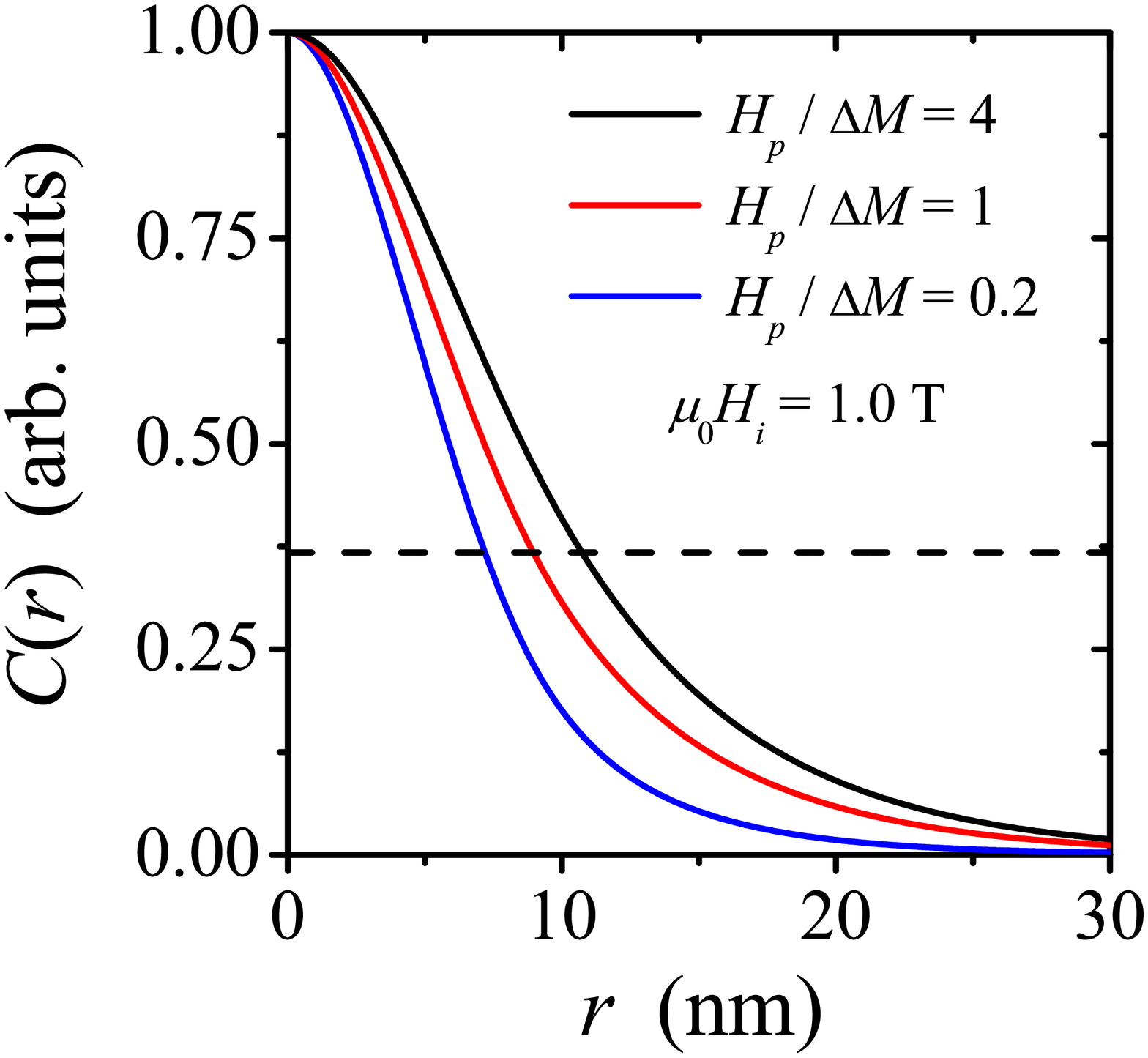}}
\caption{(Color online) Normalized correlation function $C(r)$ of the spin misalignment at $\mu_0 H_i = 1.0 \, \mathrm{T}$ and for different ratios of $H_p / \Delta M$ (decreasing from top to bottom, see inset) ($\mathbf{k}_0 \perp \mathbf{H}_0$). Dashed horizontal line: $C(r = l_C) = \exp(-1)$.}
\label{fig5}
\end{figure}

\begin{figure}[tb]
\centering
\resizebox{0.75\columnwidth}{!}{\includegraphics{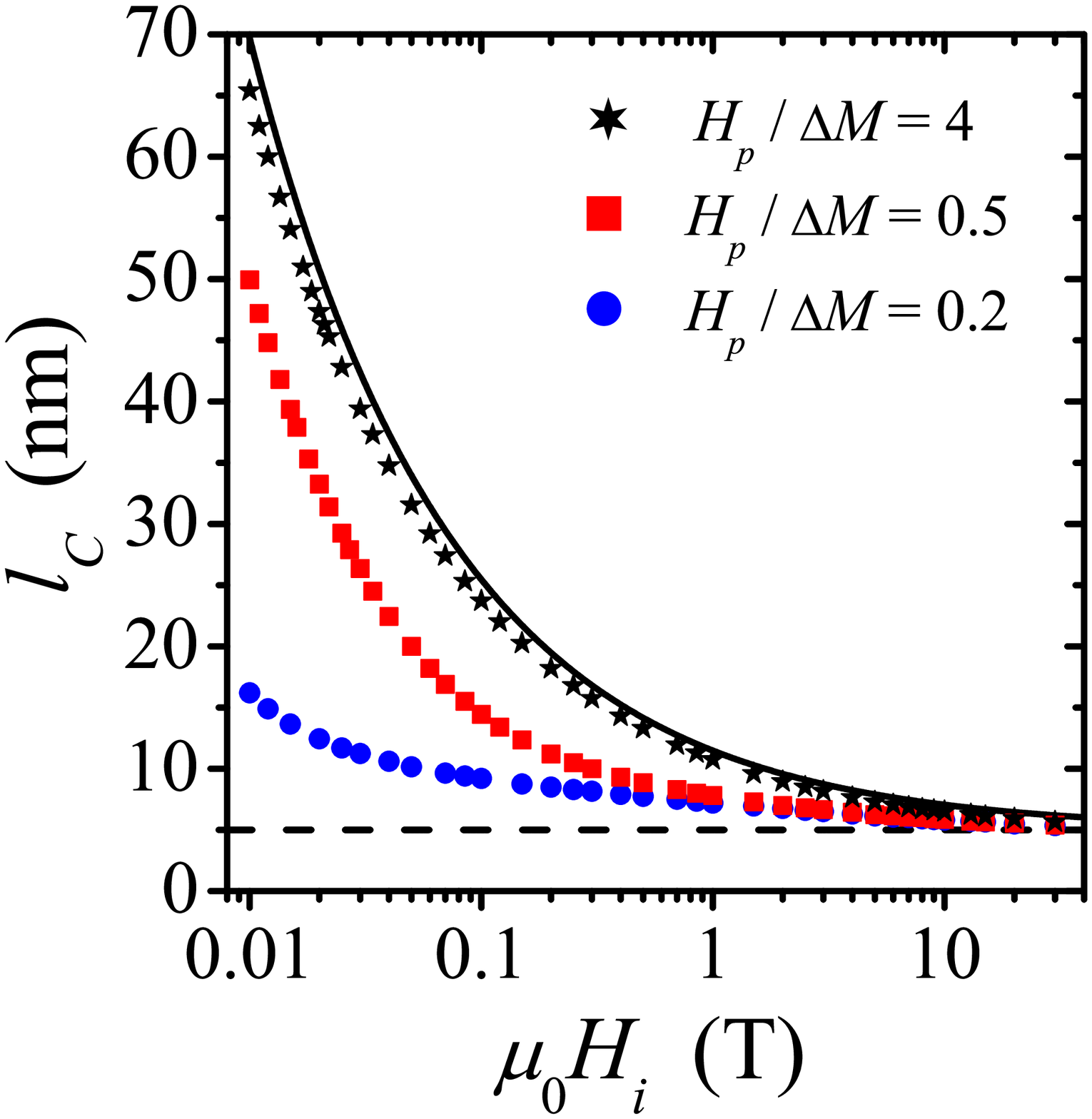}}
\caption{(Color online) Field dependence of the spin-misalignment length $l_C$ for $\mathbf{k}_0 \perp \mathbf{H}_0$ and for different ratios of $H_p / \Delta M$ (see inset) (log-linear scale). The quantity $l_C$ is identified with the $r$-value for which the correlation function has decayed to $\exp(-1)$ of its value $C(0)$ at the origin. Solid line: Eq.~(\ref{lCdef}). Dashed horizontal line: $R = 5 \, \mathrm{nm}$.}
\label{fig6}
\end{figure}

The results for $d \Sigma_M / d \Omega$ [Eq.~(\ref{sigmasmperpradav})] can be used in order to compute the autocorrelation function $C(r)$ of the spin misalignment, according to \cite{michels03prl,weissm04a,michelsprb2010}
\begin{equation}
\label{corrfunc}
C(r) \propto \frac{1}{r} \int_0^{\infty} \frac{d \Sigma_M}{d \Omega}(q) \, \sin(qr) \, q \, dq \,\,.
\end{equation}
Equation~(\ref{corrfunc}) has been solved numerically and the results for $C(r)$ at $\mu_0 H_i = 1.0 \, \mathrm{T}$ and for different ratios of $H_p / \Delta M$ are shown in Fig.~\ref{fig5}. The corresponding field dependence of the so-called correlation length $l_C$ of the spin misalignment, which quantifies the size of inhomogeneously magnetized regions, can be seen in Fig.~\ref{fig6}.

In agreement with the previous studies (e.g., Refs.~\onlinecite{michels03prl,michelsprb2010}), we find strongly field-dependent spin-misalignment correlations. The slope of the correlation function at the origin vanishes, which is in line with the absence of a sharp interface in the \emph{magnetic} microstructure. \cite{michelsprb2010,porod} The ratio $H_p / \Delta M$ decisively determines the characteristic decay length $l_C$: Increasing $H_p/\Delta M$ results in the emergence of more long-range magnetization inhomogeneities, whereas $\Delta M$ dominated perturbations in the spin structure decay on a smaller length scale. For $H_p / \Delta M \gg 1$, the relation $l_C(H_i)$ can be described by the phenomenological expression (solid line in Fig.~\ref{fig6}) \cite{lccomment}
\begin{equation}
\label{lCdef}
l_C(H_i) = R + \sqrt{\frac{2 A}{\mu_0 M_s H_i}} \,\,.
\end{equation}
Equation~(\ref{lCdef}) embodies the convolution relationship between the anisotropy-field microstructure (through the field-independent parameter $R$) and micromagnetic functions (through the field-dependent exchange length $\sqrt{2 A / (\mu_0 M_s H_i)}$. Irrespective of the value of $H_p / \Delta M$, it is seen that at large fields $l_C$ approaches the particle radius (dashed line in Fig.~\ref{fig6}).

\section{Summary and Conclusions \label{sum}}

Using the continuum theory of micromagnetics we have derived in the approach-to-saturation regime analytical expressions for the magnetic small-angle neutron scattering cross section of a two-phase particle-matrix-type ferromagnet. For the particular scattering geometry where the applied magnetic field is perpendicular to the incoming neutron beam, the results for the spin-misalignment cross section $d \Sigma_M / d \Omega$ [Eq.~(\ref{sigmasmperp})] exhibit a variety of angular anisotropies that are fundamentally different from the conventional $\sin^2\theta$ or $\cos^2\theta$-type patterns. In particular, by explicitly taking into account the wave-vector dependence of the longitudinal magnetization, novel terms appear in $d \Sigma_M / d \Omega$ which give rise to maxima roughly along the diagonals of the detector (``clover-leaf'' anisotropy), in agreement with experiment. Besides the value of the applied magnetic field, it is the ratio of the magnetic anisotropy field $H_p$ to the jump $\Delta M$ in the longitudinal magnetization at internal interfaces which determines the properties of $d \Sigma_M / d \Omega$, for instance, the asymptotic power-law exponent, the angular anisotropy, or the characteristic decay length of spin-misalignment fluctuations. Therefore, analysis of $d \Sigma_M / d \Omega$ will provide information on the strength and spatial structure of the magnetic anisotropy field, on $\Delta M$, and on the effective exchange-stiffness constant. Computation of the spin-spin correlation function corroborates the existence of long-range spin-misalignment fluctuations, with a characteristic field-dependent length that in the high-anisotropy-field limit ($H_p / \Delta M \gg 1$) can be described by Eq.~(\ref{lCdef}). By contrast, perturbations in the spin structure that are due to magnetostatic fluctuations decay on a relatively short length scale.

\begin{acknowledgments}
We thank the Deutsche Forschungsgemeinschaft (Project No.~MI 738/6-1) and the National Research Fund of Luxembourg (ATTRACT Project No.~FNR/A09/01) for financial support. Critical reading of the manuscript by Jens-Peter Bick, Frank D\"obrich and Andr$\mathrm{\acute{e}}$ Heinemann is gratefully acknowledged.
\end{acknowledgments}

\bibliographystyle{apsrev}

\begin{thebibliography}{49}
\bibitem[{lcc()}]{lccomment}
\bibinfo{note}{Note that the applicability of Eq.~(44) is essentially
  restricted to soft magnetic materials. For hard magnets, such as terbium, the
  field dependence of $l_C$ may be relatively weak and quite complicated
  (Ref.~\onlinecite{weissm04a}).}
\end{thebibliography}

\end{document}